\documentclass[aps,superscriptaddress,nofootinbib,onecolumn,a4paper,11pt]{revtex4-2}
\usepackage{amsfonts,amsmath,amssymb}
\usepackage{graphicx}
\usepackage{natbib}
\usepackage{bm}
\usepackage{hyperref}
\usepackage{dcolumn}
\usepackage{graphicx}
\usepackage{color}
\usepackage{slashed}
\usepackage[usenames,dvipsnames]{xcolor}
\usepackage{soul}
\usepackage{verbatim}
\usepackage{ulem}
\usepackage{float}
\usepackage{xcolor}

\def\mf{{\mbox{\tiny MFA}}}
\def\phiconj{\Phi^\ast}
\begin{document}

\title{Phase structure of magnetized quark matter at imaginary chemical potential}
\author{J.P.~Carlomagno}
\email{carlomagno@fisica.unlp.edu.ar}
\affiliation{Instituto de F\'isica La Plata, CONICET $-$ Departamento de F\'isica, Facultad de Ciencias Exactas,
Universidad Nacional de La PLata, C.C. 67, (1900) La Plata, Argentina}
\affiliation{CONICET, Rivadavia 1917, (1033) Buenos Aires, Argentina}
\author{D.~G\'omez~Dumm}
\affiliation{Instituto de F\'isica La Plata, CONICET $-$ Departamento de F\'isica, Facultad de Ciencias Exactas,
Universidad Nacional de La PLata, C.C. 67, (1900) La Plata, Argentina}
\affiliation{CONICET, Rivadavia 1917, (1033) Buenos Aires, Argentina}
\author{N.N.~Scoccola}
\affiliation{CONICET, Rivadavia 1917, (1033) Buenos Aires, Argentina}
\affiliation{Physics Department, Comisi\'{o}n Nacional de Energ\'{\i}a
At\'{o}mica, Avenida del Libertador 8250, 1429 Buenos Aires,
Argentina}

%%%%%%%%%%%%%%%%%%%%%%%%%%%%%%%%%%%%%%%%%%%%%%%%%%%%%%%%%%%%%%%%%%%%%%%%%%%%%%%%%%%%%%%%
\begin{abstract}
We investigate the phase structure of magnetized quark matter within a
nonlocal SU(2) Polyakov-Nambu-Jona-Lasinio model. Our work focuses on the
interplay between temperature and imaginary chemical potential in the
presence of an external uniform magnetic field, with emphasis on the
deconfinement and Roberge-Weiss transitions. We analyze the dependence of
the relevant critical temperatures on the external field, and we explore the
sensitivity of the phase structure to different Polyakov loop effective
potentials. Our results show qualitative agreement with lattice QCD
findings, reproducing inverse magnetic catalysis in the chiral sector and a
decreasing behavior of the Roberge-Weiss transition temperature with
increasing magnetic field.
\end{abstract}

%%%%%%%%%%%%%%%%%%%%%%%%%%%%%%%%%%%%%%%%%%%%%%%%%%%%%%%%%%%%%%%%%%%%%%%%%%%%%%%%%%%%%%%%
\maketitle \hfill

%%%%%%%%%%%%%%%%%%%%%%%%%%%%%%%%%%%%%%%%%%%%%%%%%%%%%%%%%%%%%%%%%%%%%%%%%%%%%%%%%%%%%%%%
\section{Introduction}

The exploration of the phase diagram of Quantum Chromodynamics (QCD) is a
central topic in modern high-energy physics. Understanding the transitions
associated with chiral symmetry restoration and the deconfinement of color
degrees of freedom is essential for describing both the evolution of the
early Universe and the properties of compact astrophysical objects. In this
context, the study of QCD matter under strong external magnetic fields has
attracted considerable attention in recent
years~\cite{Kharzeev:2012ph,Andersen:2014xxa,Miransky:2015ava,Adhikari:2024bfa}.
This interest is motivated by the fact that intense magnetic fields are
expected to arise in a variety of physical environments. For example,
magnetic fields as strong as $10^{20}\,\mathrm{G}$ may have been generated
during the electroweak phase transition in the early
Universe~\cite{Vachaspati:1991nm}. Likewise, ultra-peripheral heavy-ion
collisions are expected to produce transient magnetic fields whose strength
increases with the collision energy, reaching values of approximately
$5\times10^{19}\,\mathrm{G}$~\cite{Deng:2012pc}. In astrophysical settings,
magnetars exhibit surface magnetic fields of the order of
$10^{14}\,\mathrm{G}$~\cite{Duncan:1992hi}, while significantly stronger
fields are believed to exist in their interiors. Since these magnetic field
strengths are comparable to the square of the QCD confinement scale ($|eB|
\gtrsim \Lambda_{\rm QCD}^2$ for $|B| \gtrsim 10^{19}\,\mathrm{G}$), they
provide a unique opportunity to probe the QCD phase diagram and the
mechanisms governing deconfinement and chiral symmetry restoration.

The strongly coupled nature of QCD in this regime requires the use of
nonperturbative methods. In principle, lattice QCD (LQCD) provides a
reliable first-principle framework for studying strongly interacting matter.
However, at finite baryon density, LQCD simulations are hindered by the
well-known sign problem: when a real chemical potential is introduced, the
fermion determinant becomes complex, rendering standard Monte Carlo
importance-sampling techniques inapplicable. To circumvent this difficulty,
several alternative approaches have been developed, among which the use of
an imaginary chemical potential is particularly effective. In this case, the
fermion determinant remains real and positive, enabling direct numerical
simulations and providing valuable insights into the QCD phase structure.

Complementary insights can be obtained from effective models that capture
the essential features of low-energy QCD while remaining consistent with
available LQCD results. Among these, the Nambu--Jona-Lasinio (NJL)
model~\cite{Nambu:1961tp,Nambu:1961fr} has become one of the most widely
employed frameworks. In this model, quarks interact through local
four-fermion couplings, leading to spontaneous chiral symmetry breaking when
the interaction strength exceeds a critical
value~\cite{Vogl:1991qt,Klevansky:1992qe,Hatsuda:1994pi}. The model can be
further improved by introducing nonlocal separable interactions, which
emerge naturally in several effective approaches to QCD and generally
provide better agreement with LQCD
calculations~\cite{Schmidt:1994di,Burden:1996nh,Bowler:1994ir,Ripka:1997zb}.
A comprehensive review of nonlocal NJL models and their applications to
strongly interacting matter under extreme conditions can be found in
Ref.~\cite{Dumm:2021vop}.

The thermodynamics of QCD at finite imaginary chemical potential and
vanishing magnetic field has been extensively investigated over the last two
decades~\cite{Roberge:1986mm,deForcrand:2002hgr,Wu:2006su,deForcrand:2009zkb,DElia:2002tig,Bazavov:2017dus}.
Taking into account the presence of external magnetic fields, recent LQCD
studies of QCD at imaginary chemical potentials have revealed a rich phase
structure, including inverse magnetic catalysis effects associated with both
chiral symmetry restoration and the Roberge-Weiss (RW)
transition~\cite{Zambello:2024ucs,MarquesValois:2025nzo,DElia:2025ybj}. In
particular, LQCD simulations have provided detailed information on the
structure of the RW transition and its endpoint, as well as on the interplay
between confinement and chiral symmetry restoration in this regime. These
studies have served as important benchmarks for a variety of effective
approaches, including NJL, Polyakov-NJL (PNJL) and quark--meson models.
However, despite significant progress, neither lattice simulations nor
effective model studies have yet provided a fully conclusive determination
of the precise nature of the RW transition line away from the endpoint, in
particular whether the transition remains as a smooth crossover or it
becomes first order before reaching the critical endpoint.

In this work we investigate the phase structure of magnetized quark matter
within the framework of a nonlocal two-flavor Polyakov-Nambu-Jona-Lasinio
(nlPNJL) model~\cite{Contrera:2007wu,Hell:2008cc,Carlomagno:2013ona}. By
coupling quarks to the Polyakov
loop~\cite{tHooft:1977nqb,Polyakov:1978vu,Meisinger:1995ih,Fukushima:2003fw,Megias:2004hj,Ratti:2005jh,Roessner:2006xn,Mukherjee:2006hq,Sasaki:2006ww},
the model incorporates an order parameter for the confinement--deconfinement
transition and yields a chiral restoration temperature in good agreement
with LQCD results~\cite{Karsch:2003jg}. An important feature of nonlocal
PNJL models is their ability to naturally reproduce the inverse magnetic
catalysis (IMC) effect~\cite{Pagura:2016pwr,GomezDumm:2017iex}. In fact, the
observation of IMC in LQCD simulations~\cite{Bali:2011qj,Bali:2012zg}
represents a challenge for many conventional low-energy QCD models
(including local NJL and PNJL
approaches)~\cite{Andersen:2014xxa,Kharzeev:2012ph,Miransky:2015ava,Adhikari:2024bfa}.

The main purpose of this work is to extend previous analyses of nlPNJL model
features to a system at a finite imaginary chemical potential in the
presence of a static uniform magnetic field, thereby enabling a direct
comparison with recent LQCD results obtained under similar
conditions~\cite{MarquesValois:2025nzo,Zambello:2024ucs,DElia:2025ybj}. By
examining the interplay among temperature, imaginary chemical potential and
external fields, we aim to provide a consistent theoretical description of
the resulting phase structure and the associated phase transitions.

This paper is organized as follows. In Sec.~\ref{sec:model}, we introduce
the nlPNJL model and describe the formalism employed to incorporate both the
external magnetic field and the imaginary chemical potential. In
Sec.~\ref{sec:results}, we present and discuss our numerical results,
focusing on the behavior of the relevant order parameters and the resulting
phase structure. Finally, in Sec.~\ref{sec:conclusions} we summarize our
main findings.

%%%%%%%%%%%%%%%%%%%%%%%%%%%%%%%%%%%%%%%%%%%%%%%%%%%%%%%%%%%%%%%%%%%%%%%%%%%%%%%%%%%%%%%%
\section{Theoretical Formalism}
\label{sec:model}

We start by defining the effective Euclidean action in our nonlocal NJL
model for two quark flavors, $u$ and $d$,
\begin{equation}
S_{E}=\int d^{4}x \left[\bar{\psi}(x)\left(-i\slashed{\partial} + m_{c}\right)\psi(x)-\frac{G}{2}j_{a}(x)j_{a}(x)\right]\ ,
\label{accionE}
\end{equation}
where $\psi$ denotes the quark field doublet and $G$ is a coupling
constant. We assume that the current quark mass is the same for both
flavors, namely $m_{c}=m_{u}=m_{d}$. The nonlocal currents $j_{a}(x)$ are defined
as
\begin{equation}
j_{a}(x)=\int d^{4}z\ \mathcal{G}(z)\;\bar{\psi}\left(x + \frac{z}{2}\right)\Gamma_{a}\,\psi\left(x - \frac{z}{2}\right)\ ,
\label{corrientes}
\end{equation}
where $\Gamma_{a}=(1,i\gamma_{5}\vec{\tau})$ and $\mathcal{G}(z)$ is a form
factor. To introduce the interaction with an external magnetic field
$\vec{B}$, we replace the partial derivative $\partial_{\mu}$ in the kinetic
term of the effective action in Eq.~(\ref{accionE}) by a covariant
derivative, i.e.
\begin{equation}
\partial_{\mu} \ \rightarrow \ D_{\mu}\equiv \partial_{\mu}-i\hat{Q}\mathcal{A}_{\mu}\ ,
\end{equation}
where $\mathcal{A}_{\mu}$ stands for the external electromagnetic field, and
$\hat{Q}={\rm diag}(q_{u}, q_{d})$, with $q_{u}=2e/3$, $q_{d}=-e/3$, is the
electromagnetic quark charge operator. In order to preserve gauge
invariance, this replacement also implies a modification of the nonlocal currents
in Eq.~(\ref{corrientes}), given by
\cite{GomezDumm:2006vz,Noguera:2008cm}
\begin{gather}
\psi(x-z/2) \ \rightarrow \ \mathcal{W}(x,x-z/2)\,\psi(x-z/2)\ ,\nonumber\\
\psi(x+z/2)^{\dagger} \ \rightarrow \ \psi(x+z/2)^{\dagger}\,\mathcal{W}(x+z/2,x)\ ,
\end{gather}
where the function $\mathcal{W}(s,t)$ is defined as
\begin{equation}
\mathcal{W}(s,t)=P\,\exp\left[-i\int_{s}^{t} dr_{\mu}\hat{Q}\mathcal{A}_{\mu}(r) \right]\ .
\label{Funcw}
\end{equation}

As stated above, we consider here the case of a static and uniform
external magnetic field. We choose this field to be oriented along the
3-axis and work in the Landau gauge, taking
$\mathcal{A}_{\mu}=Bx_{1}\delta_{\mu 2}$. As is usually
done~\cite{Bloch:1952qkt}, we take a straight-line path connecting $s$ and
$t$ in Eq.~(\ref{Funcw}). Thus, the function $\mathcal{W}(s,t)$ is given by
\begin{equation}
\mathcal{W}(s,t)=\exp\left[-\frac{i}{2}\,\hat{Q}\,B\left(s_{1} + t_{1}\right)\left(t_{2} - s_{2}\right) \right]\ .
\label{Funcw2}
\end{equation}
Since quark degrees of freedom are not observed at low energies, it is
convenient to integrate out the fermionic fields, writing the action in
terms of scalar and pseudoscalar fields $\sigma(x)$ and $\vec{\pi}(x)$,
respectively. This standard procedure leads to the bosonized
action~\cite{Noguera:2008cm,GomezDumm:2010cta,GomezDumm:2017jij}
\begin{align}
S_{\rm bos} = -\ln\,\det\mathcal{D}_{x,x'} + \frac{1}{2G}\int d^{4}x
\left[\sigma(x)\sigma(x)+\vec{\pi}(x)\cdot\vec{\pi}(x) \right]\ ,
\label{Sbos}
\end{align}
where
\begin{align}
\mathcal{D}_{x,x'} = & \ \left(-i\slashed{D}_x+m_{c}\right)\delta^{(4)}(x-x')\nonumber\\
& \,+ \,\mathcal{G}(x-x')\gamma_{0}\,\mathcal{W}(x,\bar{x})\gamma_{0}\left[\sigma(\bar{x}) +
i\gamma_{5}\vec{\tau}\cdot\vec{\pi}(\bar{x})\right]\mathcal{W}(\bar{x},x')\ ,
\end{align}
with $\bar{x}=(x+x')/2$. Here a direct product to an identity matrix in
color space is understood.

We consider now the mean-field approximation (MFA), expanding the
effective action in powers of the fluctuations $\delta \sigma$ and $\delta
\pi_a$ of the mesonic fields around their vacuum expectation values.
Spontaneous chiral symmetry breaking leads to a nonzero translationally
invariant vacuum expectation value $\bar{\sigma}$ for the $\sigma(x)$ field,
while symmetry arguments imply that the mean-field values of the
pseudoscalar fields vanish, i.e., $\pi_a(x)=0$. Thus, we have
$\sigma(x)=\bar{\sigma} + \delta \sigma(x)$ and $\vec{\pi}(x)=\delta
\vec{\pi}(x)$. The MFA bosonized action per unit volume can be written as
\begin{equation}
\frac{S_{\rm bos}^{\,\mf}}{V^{(4)}}=\frac{\bar{\sigma}^{2}}{2G} - \frac{N_{c}}{V^{(4)}}
\sum_{f=u,d} {\rm tr}\,\ln\,\mathcal{D}_{x,x'}^{\mf,f}\ ,
\label{S_bos_MFA_Vol}
\end{equation}
where $N_{c}$ is the number of colors, and the trace is taken over coordinate and
Dirac spaces. The mean field operator
$\mathcal{D}_{x,x'}^{\mf,f}$ is given by
\begin{equation}
\mathcal{D}_{x,x'}^{\mf,f} =
\left(-i\slashed{\partial}_x - q_f B x_1 \gamma_2 + m_{c}\right)\delta^{(4)}(x-x')
+ \mathcal{G}(x-x')\, \bar\sigma \exp\big[i\Phi_f(x,x')\big]\ ,
\end{equation}
where $\Phi_f(x,x') = q_fB(x_2-x'_2)(x_1+x'_1)/2$ is the so-called Schwinger
phase.

Next, using the standard Matsubara formalism, we extend our analysis to a
system at finite temperature $T$ and chemical potential $\mu$. As mentioned
above, systems at finite $T$ and nonzero (real) $\mu$, under an external
uniform magnetic field, have been separately considered in previous works
(see Refs.~\cite{GomezDumm:2017iex} and \cite{Ferraris:2021vun}). The aim of
the present analysis is to investigate the combined effect of these
thermodynamical variables, considering an imaginary chemical potential. To
account for confinement/deconfinement effects, we also include the coupling
of fermions to the Polyakov loop (PL). We assume that quarks move in a
constant color background field
$\phi=ig\delta_{\mu0}G_{a}^{\mu}\lambda^{a}/2$, where $G_{a}^{\mu}$ are
color gauge fields. It is convenient to work in the so-called Polyakov
gauge, in which the matrix $\phi$ is written in a diagonal form, involving
only two independent real variables, $\phi_{3}$ and $\phi_{8}$. In this basis one has
\begin{equation}
\phi= {\rm diag}\left( \phi_r, \phi_g, \phi_b \right)=
\phi_{3}\lambda_{3}+\phi_{8}\lambda_{8}\ ,
%\mathrm{diag}\!\left(\phi_{3}+\frac{\phi_{8}}{\sqrt{3}},-\phi_{3}+\frac{\phi_{8}}{\sqrt{3}},-\frac{2\phi_{8}}{\sqrt{3}}\right),
\label{phirgb}
\end{equation}
so that the traced Polyakov loop is given by
\begin{equation}
\Phi=\frac{1}{3}\,{\rm Tr}\,e^{i\phi/T}
=\frac{1}{3}\left[e^{i(\phi_{3}+\phi_{8}/\sqrt{3})/T}
+e^{i(-\phi_{3}+\phi_{8}/\sqrt{3})/T}
+e^{-2i\phi_{8}/(\sqrt{3}T)}\right]\ .
\label{PLOP}
\end{equation}

Under the above assumptions, after some calculations we obtain the
grand canonical thermodynamic potential for a system at finite temperature
$T$ and chemical potential $\mu$ in the presence of the external magnetic
field $\vec B$. One has~\cite{Carlomagno:2023clk}
\begin{align}
\Omega_{B,T,\mu}^{\mf} = & \ \frac{\bar{\sigma}^{2}}{2G} - T\sum_{n=-\infty}^{\infty}
\sum_{c=r,g,b} \sum_{f=u,d} \frac{B_f}{2\pi} \nonumber \\
& \times \int\frac{dp_{3}}{2\pi}\,
\bigg\{\ln\left[p_\parallel^{2} + \left( M_{0,p_\parallel}^{\lambda_{f},f} \right)^{2} \right]
+ \sum_{k=1}^{\infty}\,\ln\Delta_{k,p_\parallel}^{f} \bigg\}  + \mathcal{U}(\Phi,\phiconj,T)\ ,
\label{granpotMFA}
\end{align}
where
\begin{align}
\Delta_{k,p_\parallel}^{f}=& \ \left( 2kB_f +p_\parallel^{2} + M_{k,p_\parallel}^{+,f}M_{k,p_\parallel}^{-,f}\right)^{2}
+ p_\parallel^{2}\left(M_{k,p_\parallel}^{+,f} -M_{k,p_\parallel}^{-,f}\right)^{2}\ .
\end{align}
Here, the functions $M_{k,p_\parallel}^{\pm,f}$ are given by
\begin{equation}
M_{k,p_\parallel}^{\lambda,f} = (1-\delta_{k_\lambda,-1})\,\big(m_{c} +
\bar{\sigma}\,g_{k,p_\parallel}^{\lambda,f}\big)\ ,
\label{masa_const}
\end{equation}
where
\begin{align}
g_{k,p_\parallel}^{\lambda,f} = &\ \frac{4\pi}{B_f}\left(-1\right)^{k_{\lambda}}
\int \frac{d^{2}p_{\bot}}{\left(2\pi\right)^{2}}\ g(p_{\bot}^{2} + p_\parallel^{2})
\, \exp\left(-p_{\bot}^{2}/B_f\right)L_{k_{\lambda}}(2p_{\bot}^{2}/B_f)\ ,
\label{funcg}
\end{align}
$L_n(x)$ being the Laguerre polynomials. In the above expressions we
have used several definitions. We denote by $\vec p_\bot$ a two-momentum
vector perpendicular to the magnetic field, while the two-vector $\vec
p_\parallel$ is defined as $\vec p_\parallel= (p_{3},
\omega_{n}+i\mu-\phi_{c})$, where $\omega_{n}=(2n+1)\pi T$ are the Matsubara
frequencies corresponding to fermionic modes. We recall that the color
background fields $\phi_c$, where $c=r,g,b$, can be written in terms of the
Polyakov loop parameters $\phi_3$ and $\phi_8$ according to
Eq.~(\ref{phirgb}). Notice that the grand canonical potential in
Eq.~(\ref{granpotMFA}) involves sums over color and flavor indices,
Matsubara modes $n$ and Landau levels $k$. We have also introduced the
definitions $B_f = |q_fB|$, $s_{f}={\rm sign}(q_{f}B)$, $k_{\pm}=k-1/2\pm
s_{f}/2$ and $\lambda_f = +(-)$ for $s_f = +1(-1)$. The function $g(p^2)$ in
Eq.~(\ref{funcg}) is the Fourier transform of the nonlocal form factor
${\cal G}(z)$ introduced in Eq.~(\ref{corrientes}). It is worth mentioning
that the quantity $M_{k,p_\parallel}^{\lambda,f}$ in Eq.~(\ref{masa_const})
can be interpreted as a constituent quark mass that depends on the external
magnetic field and also on the momentum, due to the nonlocal character of
the four-fermion interaction. Finally, the function
$\mathcal{U}(\Phi,\phiconj,T)$ stands for an effective Polyakov loop
potential that accounts for gluonic self-interactions. Its explicit form and
properties are discussed in Sect.~\ref{sec:UPL}.

The integral in Eq.~(\ref{granpotMFA}) is divergent and therefore has to be
regularized. We use a prescription commonly adopted in the
literature~\cite{GomezDumm:2004sr}, in which a ``free'' term is
subtracted and then added back in a regularized form, namely
\begin{equation}
\Omega_{B,T,\mu}^{\mf, \rm reg}=\Omega_{B,T,\mu}^{\mf} -
\Omega_{B,T,\mu}^{\rm free} + \Omega_{B,T,\mu}^{\rm free,reg}.
\end{equation}
Here the ``free'' contribution $\Omega_{B,T,\mu}^{\rm free}$ is
defined as the unregularized potential
$\Omega_{B,T,\mu}^{\mf}$ in Eq.~(\ref{granpotMFA}) evaluated
at $\bar{\sigma}=0$, keeping the interaction with the magnetic field and
the Polyakov loop. The corresponding regularized contribution,
$\Omega_{B,T,\mu}^{\rm free,reg}$, is given by~\cite{Menezes:2008qt}
\begin{equation}
\Omega_{B,T,\mu}^{\rm free,reg} = -\frac{N_c}{2\pi^{2}}\sum_{f}\,B_f^2\,F(x_{f})\, -
\, T\sum_{c,f}\, \frac{B_f}{2\pi}~\sum_{k=0}^\infty
\,\alpha_k \int \frac{dp}{2\pi} \ G_{k,p}^{f}(\phi_{c},\mu,T)\ ,
\end{equation}
where
\begin{eqnarray}
F(x_{f}) & = & \zeta^{'}(-1,x_{f}) + \frac{x_{f}^{2}}{4}-\frac{1}{2}\left(x_{f}^{2} - x_{f}\right)\ln x_{f}\ ,
\nonumber \\
G_{k,p}^{f}(\phi_{c},\mu,T) & =  & \sum_{s=\pm} \ln \left\{ 1 +
\exp\big[-( \epsilon_{kp}^f + s(\mu + i \phi_{c})) /T \big]\right\}\ .
\end{eqnarray}
Here we have used the definitions $x_{f}=m_{c}^{2}/(2B_f)$,
$\alpha_{k}=2-\delta_{k,0}$, $\epsilon_{kp}^{f}=\left(2kB_f +p^{2}
+m_{c}^{2}\right)^{1/2}$ and
$\zeta^{'}(-1,x_{f})=d\zeta(z,x_{f})/dz|_{z=-1}$, where $\zeta(z,x_{f})$ is
the Hurwitz zeta function.

Now, $\bar{\sigma}(B,T,\mu)$ and $\Phi(B,T,\mu)$ can be obtained by solving
the coupled set of equations that minimize $\Omega_{B,T,\mu}^{\mf, \rm reg}$, namely
\begin{align}
\frac{\partial \Omega_{B,T,\mu}^{\mf,\rm reg}}{\partial \bar{\sigma}} = 0\ ,
 & \qquad    \frac{\partial \Omega_{B,T,\mu}^{\mf,\rm reg}}{\partial \phi_3} = 0\ ,
 \qquad    \frac{\partial \Omega_{B,T,\mu}^{\mf,\rm reg}}{\partial \phi_8} = 0\ .
\label{gap}
\end{align}
Other relevant phenomenological and thermodynamic quantities can be
readily derived from the thermodynamic potential in Eq.~(\ref{granpotMFA}).
In particular, the regularized quark-antiquark condensates are given by
\begin{equation}
\langle \bar q_f q_f \rangle_{B,T,\mu}^{\rm reg}=\frac{\partial
\Omega_{B,T,\mu}^{\mf,\rm reg}}{\partial {m_{c}}}\ .
\label{average}
\end{equation}
To compare our results with those obtained from LQCD calculations~\cite{Bali:2012zg},
it is useful to define a normalized flavor-averaged condensate
\begin{equation}
 \bar \Sigma_{B,T}  = \frac{1}{2}\,(\Sigma^u_{B,T} + \Sigma^d_{B,T})\ ,
\label{qqOP}
\end{equation}
where
\begin{equation}
 \Sigma^f_{B,T} = - \frac{2 m_c}{S^4}  \left[\langle \bar q_f q_f \rangle_{B,T,\mu}^{\rm reg} - \langle \bar q_f q_f \rangle_{0,0,0}^{\rm reg}
 \right]  + \, 1\ .
\end{equation}
$S$ being a phenomenological reference scale fixed as $S = (135 \times 86)^{1/2}$~MeV.

In the crossover region, the critical temperatures associated with chiral
symmetry restoration and deconfinement transitions are defined through
the maxima of the corresponding susceptibilities. More specifically, the
chiral pseudocritical temperature $T_{\rm ch}$ is identified with the
position of the maximum of the chiral susceptibility $\chi_{\rm ch}$, while
the deconfinement pseudocritical temperature $T_{\Phi}$ is determined from
the maximum of the Polyakov loop susceptibility $\chi_{\Phi}$. These
susceptibilities are defined as
\begin{equation}
\chi_{\rm ch} =
-\frac{\partial \bar \Sigma_{B,T}}{\partial T}\ ,
\qquad\qquad
\chi_{\Phi} = \frac{\partial \Phi}{\partial T} \ .
\label{defsusc}
\end{equation}

Let us consider now the case of a nonzero imaginary chemical potential. As
is well known, Roberge and Weiss showed~\cite{Roberge:1986mm} that the
thermodynamic potential of QCD in the presence of an imaginary chemical
potential $\mu = i\, \theta\, T$ is a periodic function of $\theta$ with
period $2 \pi/3$. This implies that QCD remains invariant under a combined
transformation consisting of a $Z_3$ rotation of the quark and gauge fields
together with a shift $\theta \rightarrow \theta + 2\ell\pi/3$, where $\ell$ is an
integer. In fact, it has been demonstrated that this so-called extended
$Z_3$ symmetry is also preserved within both local and nonlocal
Polyakov-Nambu-Jona-Lasinio models~\cite{Sakai:2008um,Pagura:2011rt}. In our
framework, it can be easily seen that the thermodynamic potential is
invariant under the transformation
\begin{eqnarray}
\theta \ \rightarrow \ \theta + 2\ell\pi/3 \ , \qquad
\phi_8\ \rightarrow \ \phi_8 - 2\ell\pi T/\sqrt{3}  \ , \qquad
\phi_3\ \rightarrow \ \phi_3\ .
\label{transfsym}
\end{eqnarray}
which, according to Eq.~(\ref{PLOP}), leads to
\begin{eqnarray}
\Phi \ \rightarrow \ \exp(-i\, 2\ell\pi/3)\,\Phi \ .
\label{extz3}
\end{eqnarray}
The RW periodicity can be understood as a remnant of the $Z_3$ symmetry in
the pure gauge limit. When dynamical quarks are included in QCD and the
temperature exceeds a certain threshold $T_{RW}$, three distinct $Z_3$ vacua
emerge. These vacua are characterized by the phases of the Polyakov loop,
namely $\varphi$, $\varphi+2 \pi/3$, and $\varphi+4 \pi/3$. Roberge and
Weiss demonstrated that for $T > T_{RW}$ a first-order phase transition
occurs at $\theta=\pi/3$ modulo $2 \pi/3$, where the system jumps between
different $Z_3$-related vacua. This phenomenon is referred to as the
Roberge-Weiss transition. The endpoint of the corresponding transition line
in the $(T,\theta)$ plane, located at $(T,\theta) = (T_{RW}, \pi/3)$, is
known as the RW endpoint.

%%%%%%%%%%%%%%%%%%%%%%%%%%%%%%%%%%%%%%%%%%%%%%%%%%%%%%%%%%%%%%%%%%%%%%%%%%%%%%%%%%%%%%%%
\section{Results}
\label{sec:results}

\subsection{NJL parameters}

To obtain definite numerical results for the physical quantities of
interest, we need to specify the shape of the nonlocal form factor $g(p^2)$
in Eq.~(\ref{funcg}) and the model parameters $m_c$ and $G$. For simplicity,
we consider a Gaussian form factor
\begin{equation}
g(p^{2}) = \exp\left( -p^{2}/\Lambda^{2} \right)\ .
\end{equation}
This allows us to perform the integral in Eq.~(\ref{funcg}) analytically,
leading to a significant reduction in computation time in subsequent
numerical calculations. For this form factor the effective
constituent quark masses in Eq.~(\ref{masa_const}) are given
by~\cite{GomezDumm:2017iex}
\begin{equation}
M_{k,p_\parallel}^{\lambda,f} = (1-\delta_{k_\lambda,-1}) \left[ m_{c} \,
+\, \bar{\sigma} \frac{\left(
1-B_f/\Lambda^{2}\right)^{k_\lambda}}{\left(1+B_f/\Lambda^{2}\right)^{k_\lambda+1}}
\,\exp(-p_\parallel^{2}/\Lambda^{2})\right]\ .
\end{equation}
In any case, based on previous analyses within this type of model, we do not
expect the results to show qualitative changes when taking different form
factors shapes~\cite{GomezDumm:2017iex}. On the other hand, notice that the
form factor introduces a new dimensionful parameter $\Lambda$, which can be
understood as an effective soft cutoff momentum scale.

To fix the free parameters $m_c$, $G$ and $\Lambda$ we require the model to
reproduce the empirical values of the pion mass and decay constant,
$m_{\pi}=139$~MeV and $f_{\pi}=92.4$~MeV, at vanishing $T$, $\mu$ and $B$.
In addition, we fit the parameters to a phenomenologically acceptable value
of the chiral quark-antiquark condensate, namely $200~\mathrm{MeV} <
-\langle{\bar{q}q}\rangle^{1/3} < 300~\mathrm{MeV}$. For definiteness
we adopt here the parameter set $m_{c}=6.5$ MeV, $\Lambda=678$ MeV, and
$G\Lambda^{2}=23.66$ \cite{GomezDumm:2017iex}. This choice corresponds to a
phenomenological value of the chiral quark condensate $(-\langle \bar{q} q
\rangle)^{1/3}=230$ MeV, which provides a good agreement with LQCD
results~\cite{GomezDumm:2017iex,Carlomagno:2023clk}.

Once the parameter set has been chosen, we can solve numerically the gap
equations in Eq.~(\ref{gap}) for given values of $T$, $\mu$, and $B$. As
expected, there are regions in which, for a fixed magnetic field, more than
one solution exist for each value of $T$ and $\mu$. As usual, we consider
the stable solution to be the one corresponding to the absolute minimum of
the thermodynamic potential $\Omega_{B,T,\mu}^{\mf,\rm reg}$.

\subsection{Effective Polyakov loop potential parameters}
\label{sec:UPL}

As stated in Sect.~\ref{sec:model}, in our model the self-interactions of
color gauge fields are taken into account through an effective Polyakov loop
potential $\mathcal{U}(\Phi,\phiconj,T)$. In the following, we introduce
two forms for this potential that have been widely used in the literature.
One of them is the polynomial form~\cite{Ratti:2005jh}
\begin{equation}
\frac{\mathcal{U}_{\rm pol}(\Phi,\bar{\Phi},T)}{T^{4}}=
-\frac{b_{2}(T)}{2}\,\Phi\phiconj
-\frac{b_{3}}{6}\,\big(\Phi^3+{\phiconj}^{3}\,\big)
+\frac{b_{4}}{4}\left(\Phi\phiconj\right)^{2},
\end{equation}
where
\begin{equation}
b_{2}(T)=a_{0}+a_{1}\left(\frac{T_{0}}{T}\right)+a_{2}\left(\frac{T_{0}}{T}\right)^{2}+a_{3}\left(\frac{T_{0}}{T}\right)^{3}.
\end{equation}
Alternatively, it has been proposed to consider a function that
includes the logarithm of the Haar measure associated with the SU(3) color
group integration. This form, parametrized so as to fit pure gauge LQCD
results, is given by~\cite{Roessner:2006xn}
\begin{equation}
\frac{\mathcal{U}_{\rm log}(\Phi,\phiconj,T)}{T^{4}}=-\frac{a(T)}{2}\,\Phi\phiconj + b(T)\ln\!
\left[1-6\,\Phi\phiconj+4\,\big(\Phi^{3}+{\phiconj}^{3}\,\big)-3\,(\Phi\phiconj)^{2}\right]\ ,
\end{equation}
where
\begin{equation}
a(T)=a_{0}+a_{1}\left(\frac{T_{0}}{T}\right)+a_{2}\left(\frac{T_{0}}{T}\right)^{2}, \qquad b(T)=b_{3}\left(\frac{T_{0}}{T}\right)^{3}.
\end{equation}

In the above expressions, the potentials include a reference temperature
$T_0$. In the absence of dynamical quarks, this scale should be fixed at
270~MeV according to LQCD results. If one has two dynamical quarks, it has
been argued that this value should be reduced up to about
$T_0=210$~MeV~\cite{Schaefer:2007pw}. Notice that both the polynomial and
logarithmic potentials are invariant under the phase transformation in
Eq.~(\ref{extz3}).

To determine the free parameters associated with the logarithmic and
polynomial PL potentials we consider the results from LQCD calculations
reported in Refs.~\cite{Boyd:1996bx} and~\cite{Kaczmarek:2002mc}. In
particular, we consider the datasets quoted in Fig.~7 of
Ref.~\cite{Boyd:1996bx} and Fig.~4 of Ref.~\cite{Kaczmarek:2002mc}, which
provide complementary constraints on thermodynamic observables, allowing for
a consistent determination of the effective potential. The values of the
parameters obtained from this fit are reported in
Table~\ref{tab:potential_params}.

\begin{table}[htb]
\centering
\renewcommand{\arraystretch}{1.2}
\begin{tabular}{c|c|c}
 & \ ${\cal U}_{\rm log}$ parameters \ & \ ${\cal U}_{\rm pol}$ parameters \ \\
\hline
$a_0$ & $2.81 \pm 0.29$ & $9.11 \pm 2.92$ \\
$a_1$ & $-0.35 \pm 1.06$ & $-16.5 \pm 6.3$ \\
$a_2$ & $14.1 \pm 1.2$ & $26.1 \pm 9.3$ \\
$a_3$ & -- & $-18.5 \pm 4.0$ \\
$b_3$ & $-1.83 \pm 0.10$ & $3.0 \pm 11.9$ \\
$b_4$ & -- & $9.7 \pm 10.0$ \\
\end{tabular}
\caption{Parameters for the logarithmic and polynomial PL potentials arising
from a fit to LQCD data in Refs.~\cite{Boyd:1996bx,Kaczmarek:2002mc}.}
\label{tab:potential_params}
\end{table}

\begin{figure}[H]
\centering
\includegraphics[width=0.65\textwidth]{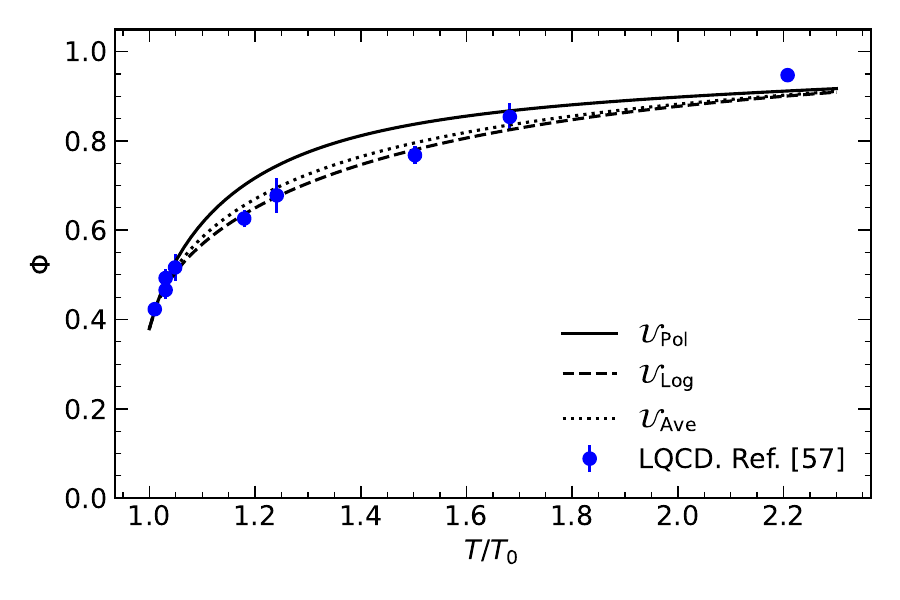}
\hfill
\hspace*{0.5cm}
\includegraphics[width=0.6\textwidth]{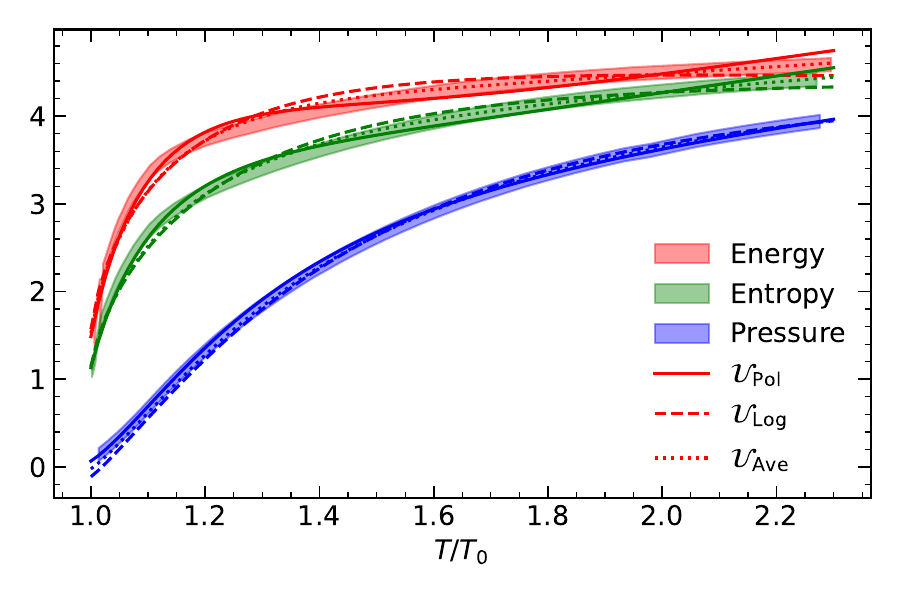}
\caption{Upper panel: Polyakov loop expectation value $\Phi$ as a function
of $T/T_0$, for the polynomial, logarithmic and average potentials. Lower
panel: energy density, entropy density and pressure as functions of $T/T_0$.
The shaded bands correspond to lattice data from Ref.~\cite{Boyd:1996bx}.}
\label{fig:PLfit}
\end{figure}

The quality of the fit is illustrated in Fig.~\ref{fig:PLfit}. In the upper
panel we show the values for the traced Polyakov loop $\Phi$ obtained from
the parameters in Table~\ref{tab:potential_params}, in comparison with
lattice data from Ref.~\cite{Kaczmarek:2002mc}. Solid and dashed lines
correspond to the polynomial and logarithmic potential, respectively. In the
lower panel we show the values obtained for three relevant thermodynamic
quantities: energy density, entropy density and pressure (red, green and
blue curves, respectively), at vanishing chemical potential and no external
field. The shaded bands correspond to LQCD results given in
Ref.~\cite{Boyd:1996bx}. In fact, the values of the parameters in
Table~\ref{tab:potential_params} are somewhat different from those quoted in
Refs.~\cite{Boyd:1996bx} and \cite{Kaczmarek:2002mc}, leading to an overall
better agreement with lattice results for the traced Polyakov loop and the
considered thermodynamical quantities. In any case, we find that these
shifts in the parameters do not imply substantial qualitative differences in
the main results of our work. Notice that the sensitivity to some of the
parameters is relatively low, which is reflected by the fact that the
corresponding entries in Table~\ref{tab:potential_params} show large errors.

Finally, in addition to the logarithmic and polynomial PL potentials, we
also consider a hybrid construction given by a linear combination of both forms,
namely
\begin{equation}
\mathcal{U}_{\rm mix}(\alpha) = \alpha\, \mathcal{U}_{\rm pol}
+ (1 - \alpha)\, \mathcal{U}_{\rm log}\ ,
\label{Umix}
\end{equation}
where $0\leq \alpha\leq 1$. The results for the traced Polyakov loop $\Phi$
and the above introduced thermodynamic quantities for the equally weighted
case $\mathcal{U}_{\rm ave}\equiv\mathcal{U}_{\rm mix}(0.5)$ are also
displayed in Fig.~\ref{fig:PLfit} (dotted lines).

\subsection{Polyakov loop potentials and phase transitions}

Before analyzing the effects of the external magnetic field on the chiral
restoration, deconfinement and RW transitions, let us first examine the
characteristics of these transitions for the above introduced forms of the
PL effective potential. To do this, we consider the mixed form in
Eq.~(\ref{Umix}), varying the mixing parameter $\alpha$ in the range from 0
to 1.

\begin{figure}[hbt]
\centering
\includegraphics[width=0.7\textwidth]{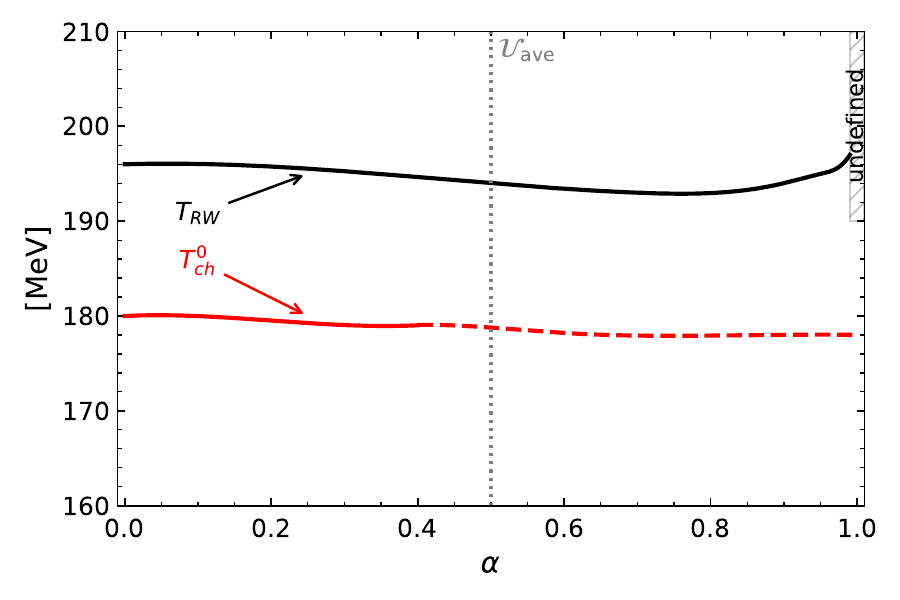}
\caption{Critical temperatures as functions of the mixing parameter
$\alpha$. Red and black lines indicate the chiral restoration temperatures
at $\mu=0$ and the RW critical temperatures, respectively. Solid (dashed)
lines correspond to first-order (crossover) transitions.}
\label{fig:Tcalpha}
\end{figure}

The curves in Fig.~\ref{fig:Tcalpha} show the dependence of the critical
temperatures $T_{\rm ch}^0$ (red) and $T_{\rm RW}$ (black) on $\alpha$ for
vanishing magnetic field. Here $T_{\rm ch}^0$ stands for the chiral
restoration temperature at vanishing chemical potential $\mu$. On one hand,
it can be seen that these critical temperatures do not show significant
changes for the full range of values of the mixing parameter. On the other
hand, it is found that the character of the chiral restoration transition
does depend on the mixing parameter. This is indicated by the line styles in
Fig.~\ref{fig:Tcalpha}, where solid and dashed lines correspond to first
order and crossover phase transitions, respectively. As shown in the figure,
for $\alpha =0$ (logarithmic PL potential) the transition is found to be of
first order, while it turns into a smooth crossover for $\alpha \gtrsim
0.4$. In addition, a singular situation is found to occur when $\alpha$ is
close to 1 (i.e., when the potential approaches the polynomial form): in
that case a RW transition temperature can be hardly defined, since, within
some finite temperature interval, three thermodynamically degenerate
solution branches coexist at the same pressure with approximately same
values of the chiral quark condensate.

Since LQCD results (for vanishing chemical potential and external fields)
are consistent with a crossover-like chiral restoration transition and a
well-defined first order RW
transition~\cite{Bornyakov:2009qh,Cuteri:2022vwk}, we conclude that the best
scenario is the one offered for a mixed PL potential, taking $\alpha$ within
a range from 0.4 up to somewhat below 1. For definiteness, in the following
calculations we consider the above introduced average potential ${\cal
U}_{\rm ave} = {\cal U}_{\rm mix}(0.5)$, which gives equal weight to the
logarithmic and polynomial PL potential forms.

\subsection{Phase transitions for nonzero external magnetic field}

We turn now to analyze the phase structure of quark matter for finite
imaginary chemical potential in the presence of an external magnetic field
$\vec B$, within the nonlocal PNJL model introduced in the previous
sections.

\begin{figure}[hbt]
\centering
\includegraphics[width=0.7\textwidth]{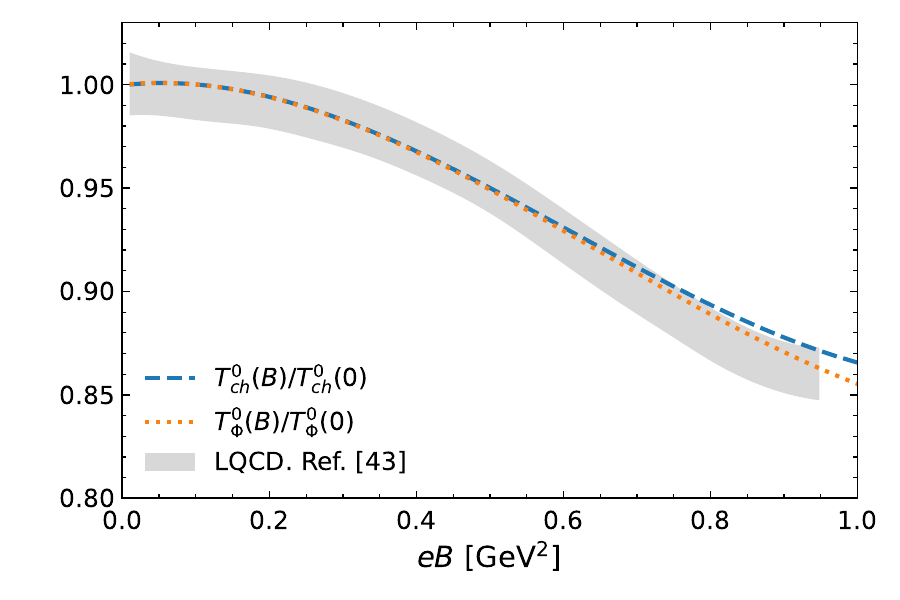}
\caption{Magnetic field dependence of the critical temperatures at
$\mu=0$, normalized to the corresponding values at $B=0$. For comparison,
LQCD results of Ref.~\cite{Bali:2012zg} are indicated by the gray band.}
\label{fig:TIMC}
\end{figure}

Firstly, we consider the restoration of chiral symmetry for the case of
vanishing chemical potential. As expected, it is seen that the values for
the critical transition temperatures decrease for increasing magnetic field,
i.e., the so-called inverse magnetic catalysis effect is observed. Our
numerical results ---corresponding to the average PL potential ${\cal
U}_{\rm ave}$--- are shown in Fig.~\ref{fig:TIMC}, where we plot the
normalized critical temperature $T_{\rm ch}^0(B)/T_{\rm ch}^0(0)$ as a
function of $eB$ (dashed line). The behavior is shown to be in good
agreement with the results obtained from lattice QCD~\cite{Bali:2012zg},
indicated by the gray band. We also show the values of the normalized
deconfinement critical temperature $T_\Phi^0(B)/T_\Phi^0(0)$, where, again,
the superindex 0 indicates that the values correspond to vanishing chemical
potential. As expected, it is found that both transitions occur
simultaneously for the studied range of values of $eB$. These findings are
also in agreement with previous results reported in
Refs.~\cite{GomezDumm:2017iex,Carlomagno:2023clk}, where the ability of
nonlocal PNJL models to reproduce IMC effects is emphasized.

Next, we consider the case of nonzero imaginary chemical potential. To
present our numerical results we introduce the normalized variable
$\eta=\theta/(\pi/3) = -i(\mu/T)(3/\pi)$. Taking into account the invariance
under the transformations in Eq.~(\ref{transfsym}), and noticing that the
thermodynamic potential is also invariant under the changes $\theta \to
-\theta$, $\phi_c\to -\phi_c$, it is seen that the studied thermodynamic
quantities are symmetric about $\eta=1$, while the first order transition
temperature at this value of $\eta$ is the above discussed Roberge-Weiss
critical temperature $T_{\rm RW}$.

\begin{figure}[hbt]
\centering
\includegraphics[width=0.49\textwidth]{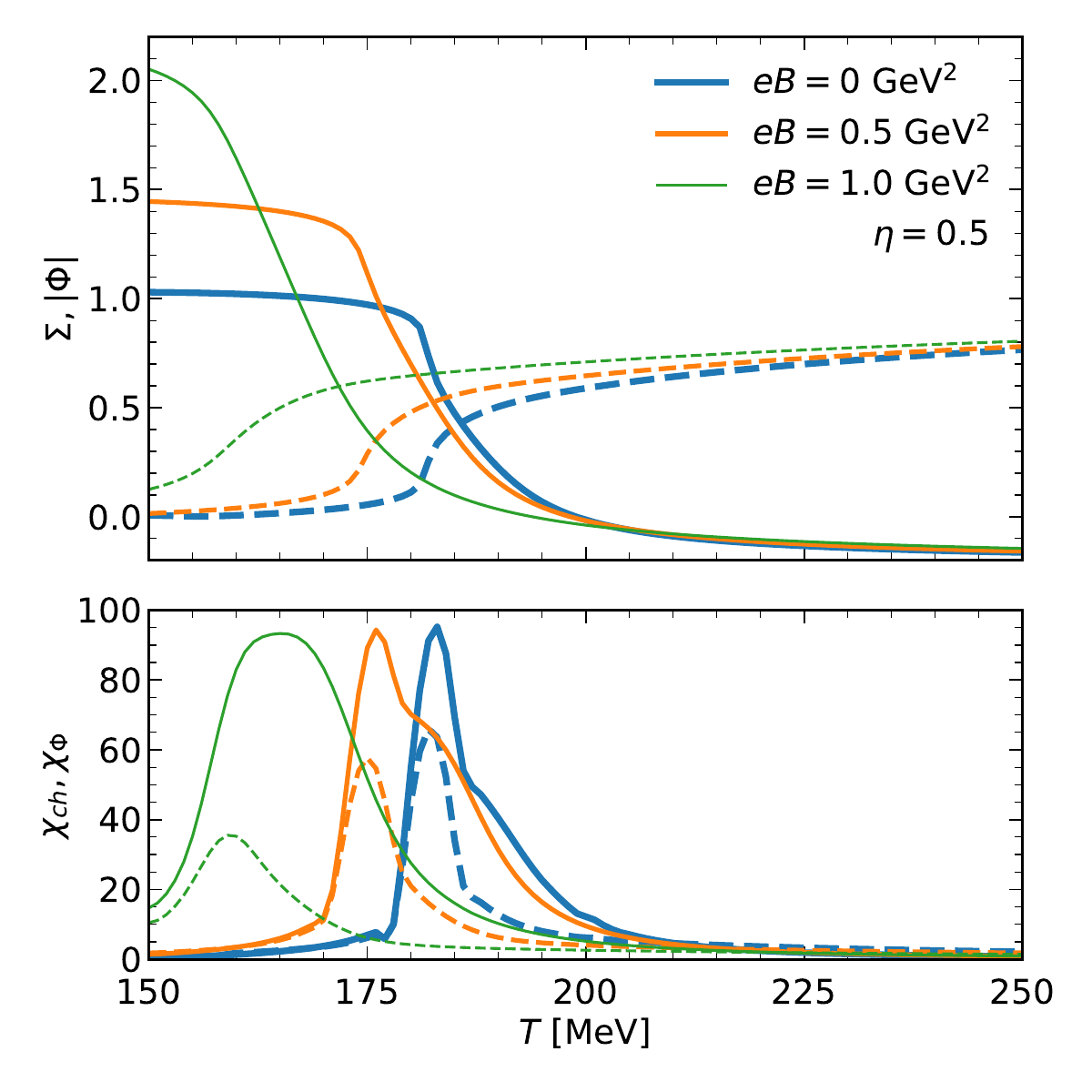}
\includegraphics[width=0.49\textwidth]{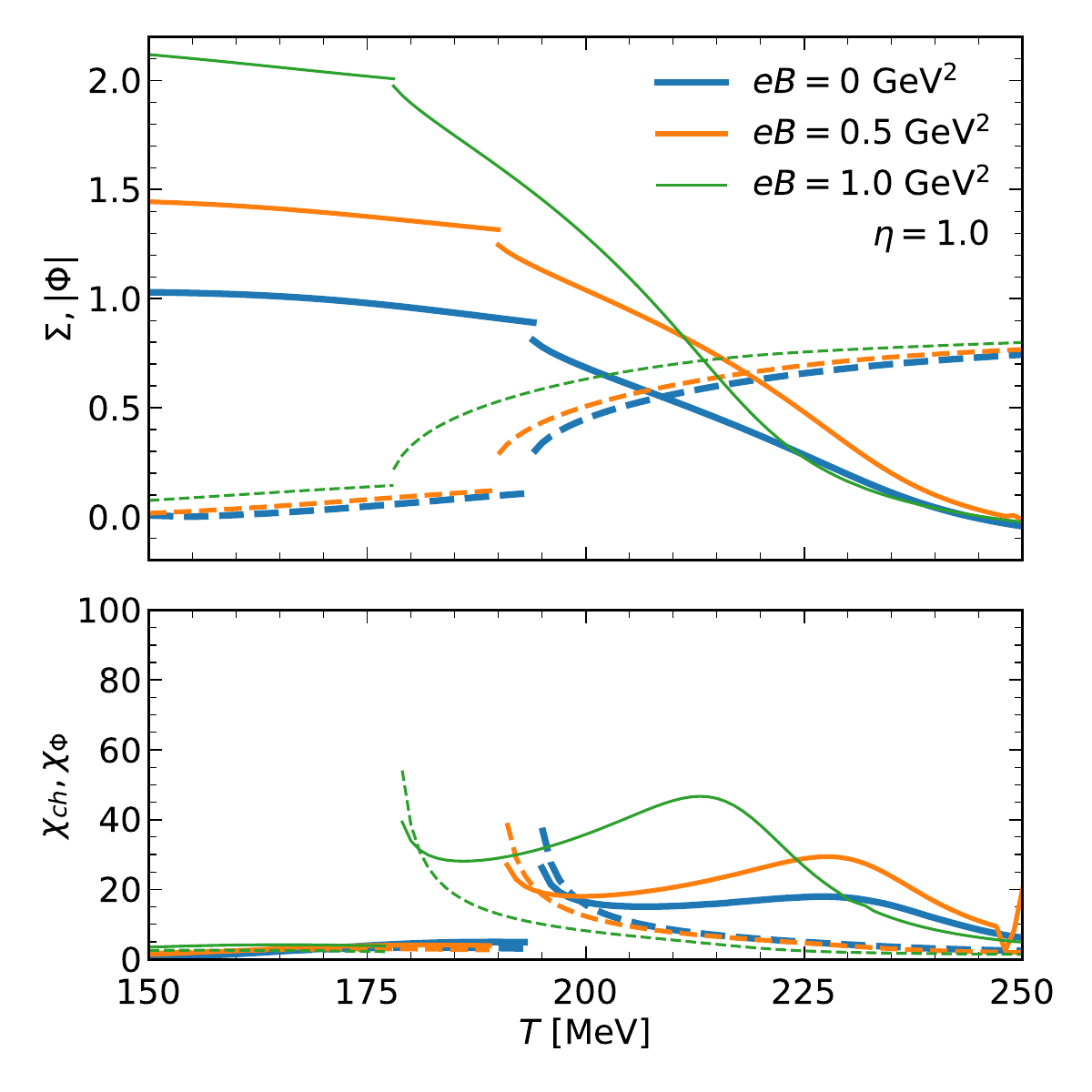}
\caption{Upper panels: temperature dependence of the order parameters
$\bar\Sigma_{B,T}$ (solid lines) and $|\Phi|$ (dashed lines), for $eB=0$,
0.5 and $1.0~\mathrm{GeV}^2$. Left and right panels correspond to $\eta =
0.5$ and $\eta=1$, respectively. Lower panels: values of the corresponding
susceptibilities as functions of $T$.}
\label{fig:SgmPhi}
\end{figure}

The thermal dependence of the order parameters for different values of the
external magnetic field are plotted in the upper panels of
Fig.~\ref{fig:SgmPhi}. Left and right panels show the results for $\eta =
0.5$ and $\eta =1$, respectively. Solid lines correspond to the values of
the averaged quark condensate $\bar\Sigma_{B,T}$, defined by
Eq.~(\ref{qqOP}), while dashed lines correspond to the absolute value of the
traced Polyakov loop, $|\Phi|$. For completeness, in the lower panels we
show the values of the associated thermal susceptibilities, according to the
definitions in Eq.~(\ref{defsusc}). For $\eta=0.5$, the position of the
peaks are understood as the critical temperatures for both deconfinement and
chiral restoration transitions, for each value of the magnetic field. On the
other hand, for $\eta = 1$ a discontinuity is found in the values of the
order parameters for all values of the magnetic field. The corresponding
temperature is associated with the above discussed Roberge-Weiss transition.
In addition, the chiral susceptibilities show a broad second peak, which can
be associated with the partial restoration of chiral symmetry. Since the
transition in this case is quite smooth, there is some ambiguity in the
determination of the corresponding critical temperatures.

\begin{figure}[hbt]
\centering
\includegraphics[width=0.7\textwidth]{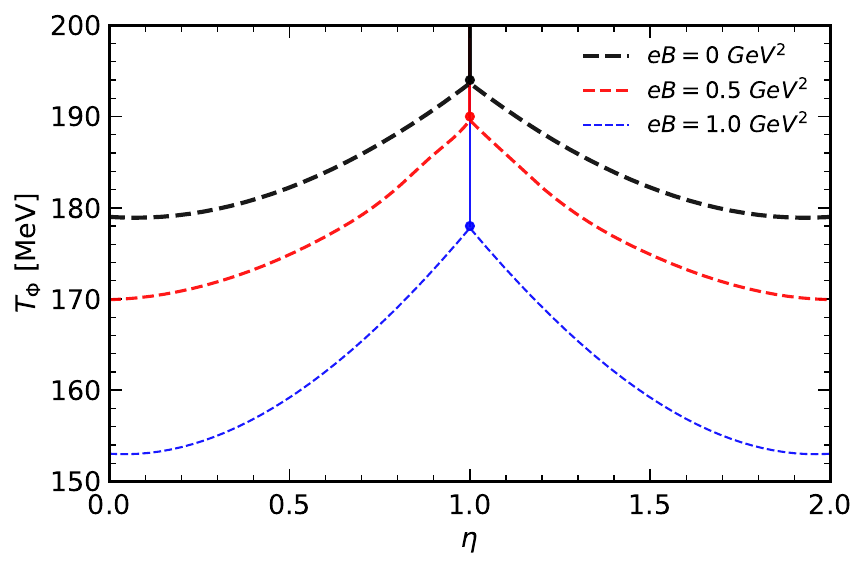}
\caption{Dependence of the critical temperature $T_\Phi$ on the
normalized parameter $\eta$ for $eB = 0$, 0.5 and $1.0$~$\mathrm{GeV}^2$.
Solid and dashed lines indicate first order and crossover-like transitions,
respectively.}
\label{fig:RWeta}
\end{figure}

In fact, we find that the deconfinement transition remains crossover-like
for all values of $\eta$ up to $\eta=1$, where the above described
discontinuity shows up. The values of the corresponding critical
temperatures $T_\Phi$ as functions of $\eta$, for different values of the
external magnetic field, are shown in Fig.~\ref{fig:RWeta}. The dashed
curves indicate that the transition lines are of crossover type for
$\eta\neq 1$, whereas at $\eta=1$ they end up at first-order RW endpoints
(fat dots in the figure). As discussed above, this $T\!-\!\eta$ phase
diagram exhibits a periodic repetition of Roberge--Weiss (RW) first-order
transition lines, which are symmetric about $\eta=1$.

Our results are found to be in qualitative agreement with the discussion
presented in Ref.~\cite{Bonati:2016pwz} for $B=0$, where a similar structure
of the phase diagram, in the context of lattice QCD, is reported: the RW
transition line is shown to remain crossover-like for $\eta<1$, becoming
genuinely of first order only at the RW endpoint. It is worth noticing that
other studies in the literature report a more intricate structure for the RW
transition line, in which a first-order segment may extend away from the
endpoint for certain choices of model parameters.

\begin{figure}[hbt]
\centering
\includegraphics[width=0.7\textwidth]{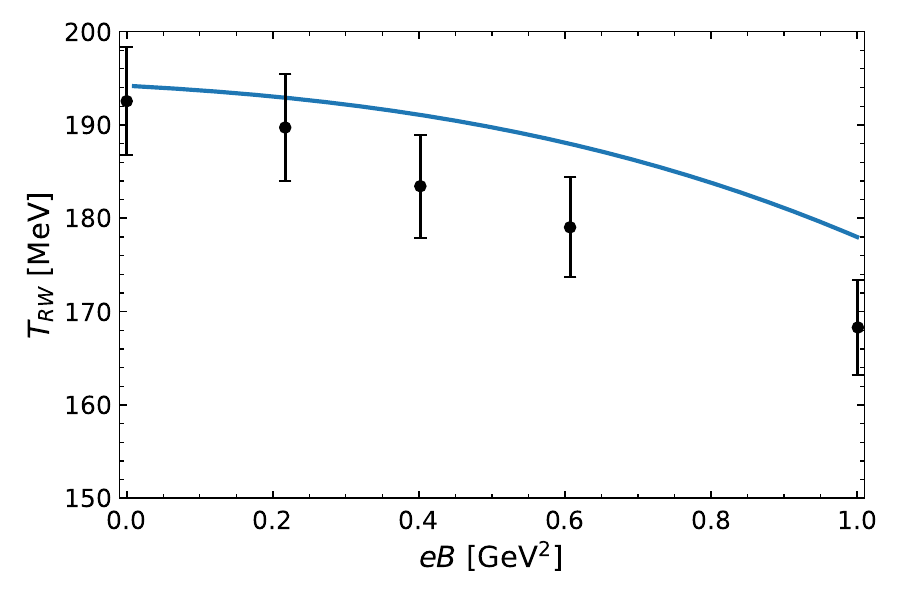}
\caption{RW critical temperature as a function of $eB$ within the
nlPNJL model considered in this work. The black dots correspond to LQCD
results from Ref.~\cite{DElia:2025ybj}.} \label{fig:TRW}
\end{figure}

Finally, in Fig.~\ref{fig:TRW} we show the dependence of the Roberge-Weiss
transition temperature $T_{\text{RW}}$ on the background magnetic field
strength. Our results are compared with recent LQCD calculations reported in
Ref.~\cite{DElia:2025ybj} (black markers). The associated error bands
indicate a 3\% uncertainty, as quoted in that work. It is seen that our
model shows a good qualitative agreement with LQCD results, reproducing the
monotonic decrease of the RW temperatures when the magnetic field is
increased. It can be said that this behavior reveals the presence of inverse
magnetic catalysis in the RW critical temperature.

We have also checked the stability of our results under moderate
variations of the PL potential parameter $\alpha$ (see Eq.~(\ref{Umix})).
While, as stated, the above results correspond to $\alpha = 0.5$, no
qualitative changes arise in either the behavior of the critical
temperatures or the overall structure of the phase transitions if this value
is increased e.g.\ up to 0.75. Only minor quantitative shifts are observed,
indicating that our main conclusions do not depend on a fine-tuned choice of
this parameter.

%%%%%%%%%%%%%%%%%%%%%%%%%%%%%%%%%%%%%%%%%%%%%%%%%%%%%%%%%%%%%%%%%%%%%%%%%%%%%%%%%%%%%%%%

\section{Conclusions}
\label{sec:conclusions}

In this work we investigate the phase structure of magnetized quark matter
within a nonlocal two-flavor Polyakov-Nambu-Jona-Lasinio model, considering
the interplay between temperature, imaginary chemical potential and external
magnetic field strength. The analysis is carried out using a mixed
Polyakov-loop potential that interpolates between logarithmic and polynomial
forms.

The dependence of the chiral restoration and deconfinement critical
temperatures on the interpolating parameter $\alpha$ is examined. While the
absolute values of the transition temperatures exhibit only a mild
sensitivity to $\alpha$, its influence becomes relevant for the nature of
the chiral restoration transition, which shifts from first-order to
crossover. In order to remain within a regime compatible with the
qualitative behavior observed in lattice QCD studies, we choose an
intermediate value $\alpha = 0.5$. It is found that at vanishing chemical
potential the model reproduces inverse magnetic catalysis, in agreement with
previous findings from nonlocal PNJL approaches and lattice QCD simulations.

For the case of a finite imaginary chemical potential, the thermal behavior
of the order parameters and their susceptibilities are calculated,
considering different values of the external magnetic field. From the
analysis of these curves, the associated phase diagram in the $T-\eta$ plane
(where $\eta = -i(\mu/T)(3/\pi)$) is constructed. In particular, for the
studied range of values of $B$ the Roberge-Weiss transition is found to be
of first order, while the corresponding critical temperatures $T_{RW}$ show
a decreasing trend with increasing magnetic field. Both results turn out to
be qualitatively consistent with those obtained in recent lattice QCD
analyses.

We conclude by mentioning that a natural continuation of the present
analysis would be the inclusion of additional quark flavors, in particular
the strange sector within an SU(3) flavor symmetry framework. In this
way one could assess the role of flavor mixing and strangeness dynamics
under strong magnetic fields and imaginary chemical potential.

%%%%%%%%%%%%%%%%%%%%%%%%%%%%%%%%%%%%%%%%%%%%%%%%%%%%%%%%%%%%%%%%%%%%%%%%%%%%%%%%%%%%%%%%
\section*{Acknowledgements}

This work has been supported in part by Consejo Nacional de Investigaciones
Cient\'ificas y T\'ecnicas (Argentina) under Grant No.~PIP2022-GI-11220210100150CO and by the National University of La Plata (Argentina), Project No.~X960.

%%%%%%%%%%%%%%%%%%%%%%%%%%%%%%%%%%%%%%%%%%%%%%%%%%%%%%%%%%%%%%%%%%%%%%%%%%%%%%%%%%%%%%%%
\bibliographystyle{apsrev4-2}
\bibliography{refs}

\end{document}